\documentclass[preprint]{aastex}

\begin{document}


\title{Formalism for inclusion of measured reaction cross sections in stellar rates including uncertainties
and its application to neutron capture in the s-process}

\author{Thomas Rauscher}
\affil{Department of Physics, University of Basel, CH-4056 Basel, Switzerland}

\begin{abstract}
A general formalism to include experimental reaction cross sections into calculations of stellar rates is presented. It also allows to assess the maximally possible reduction of uncertainties in the stellar rates by experiments. As an example for the application of the procedure, stellar neutron capture reactivities from \cite{kadonis} are revised and the remaining uncertainties shown. Many of the uncertainties in the stellar rates are larger than those obtained experimentally. This has important consequences for s-process models and the interpretation of meteoritic data because it allows the rates of some reactions to vary within a larger range than previously assumed.
\end{abstract}


\section{Introduction}
\label{sec:intro}

An increasing number of reaction cross section measurements are performed with the aim to improve reaction rates for nucleosynthesis studies. The questions of how to assess the impact of the experimentally obtained cross sections on astrophysical reaction rates and how to convert laboratory cross sections to stellar rates inevitably arise. Closely related is the question concerning the estimate of remaining uncertainties in the stellar rates. These questions are addressed in the following.

The stellar enhancement factor (SEF) -- the (theoretically predicted) ratio of stellar and laboratory rate -- was used in the past to derive the stellar rate from a measurement. It was shown in \citet{xfactor} that the SEF is not adequate for this purpose and the ground state contribution $X_0$ was introduced to replace the SEF.

Here, a further generalization of this concept is presented and a formalism for including data and their uncertainties into an improved stellar rate with revised uncertainties is laid out. The generalized relative level contribution to the stellar rate is introduced in \S~\ref{sec:The-ground-state}. The proper inclusion of data into stellar rates is discussed in
\S~\ref{sec:Renormalization-of-theory}. The following \S~\ref{sec:Determining-uncertainty-factors} then explains how
the attached error changes when using experimental information to improve a stellar rate. As an important application, neutron capture rates for the s-process are revised in \S~\ref{sec:application}. They are commonly believed to be strongly constrained by precise data but it will be shown that the remaining uncertainties are not dominated by the experimental errors in many cases.

\section{The relative contribution $X_i$ to the stellar rate\label{sec:The-ground-state}}

The stellar rate $r^{*}$ is a weighted sum over reactivities $R_i$
of thermally excited states, which are each bombarded by Maxwell-Boltzmann
energy-distributed projectiles,
\begin{equation}
r^{*}=n_{\mathrm{pro}}n_{\mathrm{tar}}\sum_{i}w_{i}R_{i}=n_{\mathrm{pro}}n_{\mathrm{tar}}R^{*}\quad,\label{eq:stellrate}
\end{equation}
with the number densities $n_{\mathrm{pro}}$, $n_{\mathrm{tar}}$
of projectile and target nucleus, respectively. The statistical weights
are given by \citep{fow74}
\begin{equation}
w_{i}=\frac{\left(2J_{i}+1\right)\exp\left(-E^\mathrm{exc}_{i}/(kT)\right)}{\sum_{m}\left\{ \left(2J_{m}+1\right)\exp\left(-E^\mathrm{exc}_{m}/(kT)\right)\right\} }=\frac{\left(2J_{i}+1\right)\exp\left(-E^\mathrm{exc}_{i}/(kT)\right)}{G}\quad,
\end{equation}
where $J_{i}$ and $E^\mathrm{exc}_{i}$ are spin and energy of the $i$th target
level and $G$ is the nuclear partition function, summing over all
excited target levels. Each reactivity is derived by folding the reaction
cross section $\sigma_i$ with the energy distribution $\Phi$ of the projectiles,
\begin{equation}
R_{i}=\int_0^\infty \sigma_i(E_i) \Phi(E_i,T)dE_i\quad,
\end{equation}
where in the case of interacting nuclei at usual stellar plasma conditions
$\Phi$ is the Maxwell-Boltzmann distribution.

The quantity
\begin{equation}
\label{eq:xfactor}
X_i(T)=\frac{w_iR_i(T)}{R^{*}(T)}=\frac{2J_i+1}{2J_0+1}e^{-E_i^\mathrm{exc}/(kT)}\frac{\int\sigma_i(E)\Phi(E,T)dE}{\int\sigma^{\mathrm{eff}}(E)\Phi(E,T)dE}
\end{equation}
describes the contribution of transitions commencing on level $i$ of a target nucleus to the \emph{stellar} rate \citep{xfactor,sensi}. The latter
also includes additional transitions originating from further thermally populated excited
states. The reaction cross section $\sigma_i$ is defined as
\begin{equation}
\label{eq:statecs}
\sigma_i=\sum_{j}\sigma^{i\rightarrow j}
\end{equation}
and the effective cross section as \citep{fow74,holadndt}
\begin{equation}
\sigma^{\mathrm{eff}}=\sum_{i}\sum_{j}\sigma^{i\rightarrow j}\quad,
\end{equation}
where $\sigma^{i\rightarrow j}$ is the partial cross section from
target level $i$ to final level $j$ (cross sections at zero or negative energies are zero).

The relative contribution $X_i$ can be used to determine the impact of laboratory
measurements on stellar rates. Laboratory cross sections only consider reactions on nuclei in a single level $i=i^\mathrm{lab}$. Usually, this is the
ground state (g.s.) of the nucleus, which implies
$i^\mathrm{lab}=0$ (when counting of the levels starts with 0 at the g.s.) and thus $\sigma^\mathrm{lab}=\sum_j \sigma^{0\rightarrow j}$. A few nuclides naturally occur in an isomeric state above the g.s. In this case, the laboratory cross section is derived using the isomeric target state $i_\mathrm{iso}$ and thus $i^\mathrm{lab}=i_\mathrm{iso}$. Among s-process nuclides there is the isomer $^{180\mathrm{m}}$Ta which is a level at $E^\mathrm{exc}_{i_\mathrm{iso}}=77$ keV excitation energy and with spin $J_{i_\mathrm{iso}}=9$.

Often used in experimental s-process studies,
the MACS (Maxwellian Averaged Cross Section) $\left\langle \sigma\right\rangle $ is nothing
else than the laboratory reactivity $R_{i^\mathrm{lab}}$ divided by the most probable projectile
energy,
\begin{equation}
\left\langle \sigma\right\rangle (T)=\frac{R_{i^\mathrm{lab}}}{kT}\quad,
\end{equation}
where $E=kT$ is the most probable energy in the Maxwell-Boltzmann energy
distribution. For use in nucleosynthesis calculations, a MACS always
has to be converted to a rate. Obviously, this cannot be a stellar rate $r^*$ when $X_i<1$. Therefore a rate directly computed from
a (measured) MACS is only useful when $X\approx1$. Otherwise, the rate
has to be calculated from theory, perhaps using some information from
the MACS, as explained in the following.

\section{Renormalization of theory to experiment\label{sec:Renormalization-of-theory}}

\subsection{General}

A theoretical \emph{stellar} rate $r_{\mathrm{th}}^{*}$ cannot be
directly renormalized to an experimental value because the measurement
only yields a rate on a single level $r_{\mathrm{exp}}^{i^\mathrm{lab}}$ or MACS $\left\langle \sigma\right\rangle _{\mathrm{exp}}$.
Therefore, only the calculated rate $r_{\mathrm{th}}^{i^\mathrm{lab}}$ on level $i^\mathrm{lab}$ or
MACS $\left\langle \sigma\right\rangle _{\mathrm{th}}$ can be compared
to the measured one, giving a ratio
\begin{equation}
v'=\frac{r_{\mathrm{exp}}^{i^\mathrm{lab}}}{r_{\mathrm{th}}^{i^\mathrm{lab}}} =
\frac{\left\langle \sigma\right\rangle _{\mathrm{exp}}}{\left\langle \sigma\right\rangle _{\mathrm{th}}}\quad,
\end{equation}
respectively. But what is actually needed for a renormalization of the stellar rate is the ratio
\begin{equation}
{v^{*}}'=\frac{r_{\mathrm{new}}^{*}}{r^{*}} \quad,
\end{equation}
where $r_{\mathrm{new}}^{*}$ is the new \textit{stellar} rate including the experimental information.
There are two possible approaches
for implementing the experimental results into a stellar quantity:
\begin{enumerate}
\item Assume that the transitions from the excited states of the target
have the same deviation as the one found for the transitions from the laboratory level $i^\mathrm{lab}$;
then ${v^{*}}'=v'$.
\item Realize that the measurement of the laboratory level transitions did not
provide any information about the transitions originating from other excited
states of the target. Therefore only the laboratory
level contribution to the stellar rate should be renormalized and the other contributions left
unchanged. This leads to a renormalization factor
\begin{equation}
{v^{*}}'=v'X_{i^\mathrm{lab}}+\left(1-X_{i^\mathrm{lab}}\right)=1+X_{i^\mathrm{lab}}\left(v'-1\right) \quad,\label{eq:renormstellar}
\end{equation}
 with $X_{i^\mathrm{lab}}$ being the laboratory contribution as before.
\end{enumerate}
Note that the renormalization factors are temperature-dependent in
both approaches.

Assumption 1 can be used when the reason for the deviation of theory
from experiment is known and also that it applies (with the same magnitude)
to the cross sections of the excited levels. This cannot be inferred
generally but would have to be investigated for each nucleus and reaction
separately. Often the errors in predicting the transitions from the g.s.\ and from excited levels may
be correlated. Nevertheless, it is not easily justified to claim that
they are the same. Assumption 2 presents a more conservative
view to the other extreme, implying that the measurement did not constrain
the cross sections of the other levels in any way and that therefore
these should not be included in a renormalization. Only the
transitions on the laboratory level are renormalized in that picture.

To be sure, using assumption 2 combined with the error estimate described
in \S~\ref{sec:Determining-uncertainty-factors} below is the safest
way to perform a renormalization.

\subsection{Selection of renormalization temperature}
\label{sec:haufeshrenorm}

It is also important to consider the applicability of the theoretical
model to calculate the cross sections or reaction rates. For compound
nuclei with low level density at the compound formation energy (for
example, nuclei at magic neutron numbers), the statistical Hauser-Feshbach
model may not be applicable \citep{rtk97}. If this is the case, then also the energy
dependence will not be correct. The latter is also true when the model
is applicable at the higher energy part of the considered (n,$\gamma$)
energies but not at lower energies.

It would be best to use renormalized Hauser-Feshbach
values only in the region where the model is applicable and employ
some other temperature/energy dependence outside that region. Unfortunately
this is only possible when there are experimental data at the low
energies, which often is not the case. If experimental data were available,
a renormalization would not be required, anyway.

When renormalizing, it has to be taken care whether the renormalization value really is at an energy or stellar temperature in the applicability range of the model. If not, a value at higher energy or temperature has to be used. Again, this
is only possible when experimental data are available not only at one energy or temperature.

\section{Determining uncertainty factors and error bars on stellar rates\label{sec:Determining-uncertainty-factors}}

Let us assume the theoretical stellar rate $r^{*}$ (or stellar reactivity)
has an uncertainty factor $U^*=U_{\mathrm{th}} \geq 1$. For example, a 50\%
uncertainty translates into $U_{\mathrm{th}}=1.5$ and the true rate
$r_{\mathrm{true}}^{*}$ is expected to be in the range $r^{*}/1.5\leq r_{\mathrm{true}}^{*}\leq1.5r^{*}$.
Let us further assume that an experiment has measured the MACS or
g.s.\ rate ($i^\mathrm{lab}=0$) with an uncertainty factor $U_{\mathrm{exp}} \geq 1$ which is
smaller than $U_{\mathrm{th}}$. When the theoretical MACS is replaced
by its experimental value, this will result in a new \emph{stellar}
rate as described in \S~\ref{sec:Renormalization-of-theory}. The
new uncertainty in this stellar value will be \citep{sensi}
\begin{equation}
U^*_\mathrm{new}=U_{\mathrm{exp}}+(U_{\mathrm{th}}-U_{\mathrm{exp}})(1-X_{i^\mathrm{lab}})\quad.\label{eq:uncertainty}
\end{equation}
Since $X_{i^\mathrm{lab}}$ depends on the plasma temperature, also the uncertainty
factors are temperature dependent.

For historical reasons, experimental uncertainties are usually given
with linear error bars instead of uncertainty factors, although
the latter would be more appropriate, especially for values approaching zero \citep[such factors imply the use of lognormally distributed errors instead of normal distributions, see, e.g.,][]{lognormal}.
They can be mapped to uncertainty
factors but a symmetric error $\pm\Delta$ (with $\Delta \geq 0$) of a value $v$ (either a
MACS or a reactivity) will lead to an asymmetric uncertainty
\begin{eqnarray}
U^{\mathrm{up}} &  & =\frac{v+\Delta}{v}\quad,\nonumber \\
U^{\mathrm{down}} &  & =\frac{v}{v-\Delta}\quad.\label{eq:updown}
\end{eqnarray}
In our definition uncertainty factors $U\geq1$ and therefore the
lower limit of $v$ is given by $v/U^{\mathrm{down}}$ and the upper
limit is given by $vU^{\mathrm{up}}$. For symmetric uncertainty factors,
as usually encountered in astrophysical rates, $U=U^{\mathrm{up}}=U^{\mathrm{down}}$
and $v$ is in the range
\begin{equation}
v/U\leq v\leq vU\quad,\label{eq:uncrange}
\end{equation}
as used above.

It is important to properly interpret the uncertainty factors defined above. Compatible definitions of theoretical and experimental uncertainties have to be used especially in Equation (\ref{eq:uncertainty}). While experimental uncertainties are usually quoted as confidence intervals from statistical distributions, theoretical errors cannot be rigorously defined in such a manner \citep[see, e.g.,][for a more detailed discussion]{sensi}. Often, theoretical upper and lower limits are given with the implicit assumption of a uniform probability for the actual value to be located anywhere within the limits. Moreover, experimental uncertainties contain statistical and systematical errors, which have to be treated differently. From the point of view of astrophysical application, i.e., variation of reaction rates, the range of values given by the uncertainty factors as shown in Equations (\ref{eq:uncertainty}) and (\ref{eq:uncrange}) should cover a reasonable range of possible reaction rate values. In varying the rates, uniform probability is usually assumed within the given range, although Monte Carlo studies also allow for normal and lognormal distributions \citep{smith02,smith04}. For practical purposes -- although mathematically not rigorous -- a simple but reasonable choice would be to assume a 2-sigma confidence interval for the experimental uncertainties $U_\mathrm{exp}$ and estimated upper and lower limits for theory values also containing 95\% of the possible values, and then to assume that the actual values are uniformly distributed across both ranges. Using a 3-sigma interval of the (non-uniformly distributed) experimental data would be, in my opinion, exaggerated because this would include outliers with low probability which are then assigned too high importance in the conversion to a uniform distribution.

Since $X_i$ is a theoretical quantity, it also has an inherent error.
The impact of its uncertainty, however,
is small with respect to the experimental and theoretical uncertainties
in cross sections and rates. It was shown in \citet{xfactor} that the magnitude of the error scales
inversely proportionally with the value of $X_i$, i.e., $X_i=1$ has
zero error\footnote{as long as $G$ is known; this is the case close to stability},
and that the uncertainty factor $U_X\geq 1$ of $X_i$ is given by $\max(u_X,1/u_X)$, where
\begin{equation}
u_X=\overline{u}\left(1-X_i\right) + X_i
\end{equation}
and $\overline{u}$ is an averaged uncertainty factor in the predicted \textit{ratios} of the $R_i$.
In any case, the uncertainties are sufficiently small to preserve the magnitude
of $X_i$, i.e., small $X_i$ remain small within errors and large $X_i$ remain large. The uncertainty factors $U_X$ can easily
be inserted in Equations (\ref{eq:renormstellar}) and (\ref{eq:uncertainty}) and the uncertainties quoted for the reactivities in
\S~\ref{sec:application} and Table \ref{tab:results} also contain $U_X$.

\section{Application to neutron capture reactions in KADoNiS}
\label{sec:application}

The \citet{kadonis} compilation \citep[see also][]{dillkadonis} gives recommended MACS for neutron capture
at 30 keV from experiment and their uncertainties. It also provides
(n,$\gamma$) reactivities in the range $5\leq kT\leq100$
keV, derived either from experiment or from a combination of experiment
and theory. Those reactivities did not consider the influence of $X_{i^\mathrm{lab}}$
and therefore have to be revised. At the same time, updated error
estimates can be provided.

For cases of rates experimentally provided over
the full temperature range, it is obvious that these are
laboratory reactivities $R_{i^\mathrm{lab}}$, as discussed above. They are equal to
the stellar reactivities $R^{*}$ as long as $X_{i^\mathrm{lab}}=1$. When $X_{i^\mathrm{lab}}<1$,
a different approach has to be taken. For each temperature in the
$5\leq kT\leq100$ keV range, the \emph{theoretical} \emph{stellar}
reactivity has to be renormalized, using Equation (\ref{eq:renormstellar}).
Each value will then have an uncertainty according to Equation (\ref{eq:uncertainty}).

The cases treated by a combination of experiment and theory were derived
by renormalizing a theory value to an experimental value at a single
temperature. The values at all other temperatures were then taken from the
renormalized theory. A similar approach can be kept when including
the influence of $X_{i^\mathrm{lab}}$. The theoretical stellar rate has to be renormalized
according to Equation (\ref{eq:renormstellar}) at one temperature, then values at all
other temperatures are given by the renormalized stellar rate. Since
the experimental uncertainty is only given at one temperature, Equation
(\ref{eq:uncertainty}) can be straightforwardly applied only when assuming
that the experimental uncertainty factors would be the same at other
energies%
\footnote{A very conservative approach -- not adopted here -- would include an experimental uncertainty
factor only in the value for the measured energy/temperature and use
pure theory uncertainties for all others.%
}.

What are the uncertainty factors $U_{\mathrm{exp}}$ and $U_{\mathrm{th}}$
to be used in Equation (\ref{eq:uncertainty})? The experimental uncertainty
is given for the recommended MACS at $kT=30$ keV in \citet{kadonis}. It is
stated in the compilation that the relative errors are expected to
be similar at all other temperatures. 
As it is common in experimental nuclear physics, the quoted errors are on the 1-sigma level. Following the procedure recommended in \S~\ref{sec:Determining-uncertainty-factors}, a 2-sigma error is used in the following to define an uncertainty range covering most of the allowed values. It is derived by doubling the error given in \citet{kadonis}, and converted to $U_\mathrm{exp}$ by applying Equation (\ref{eq:updown}). Since systematical errors are not given separately, they cannot be treated differently.

The inherent theoretical uncertainty is difficult to estimate, as
explained in \S~\ref{sec:Determining-uncertainty-factors}. For neutron capture at stability there
is a comparison to the compilation of \citet{bao} \citep[a predecessor of][]{kadonis} in \citet{rtk97}.
An \emph{average} deviation of 30\% is found, this would translate
into an uncertainty factor $U_{\mathrm{th}}=1.3$. Comparing the MACS of \citet{kadonis} to the theoretical (g.s.) MACS calculated with the code SMARAGD, v0.8.4 \citep{smaragd,raureview}, an average deviation of 10\% is found. On
the other hand, local deviations up to a factor of 2.4 are present (but this may be due to a low level density, rendering the Hauser-Feshbach model inadequate).
Moreover, the comparison is only made for MACS and not for stellar
rates. If one wants to make sure that the uncertainty range of Equation
(\ref{eq:uncrange}) really covers the possible stellar values, it may
be safer to assume $U_{\mathrm{th}}=2$. This is equivalent to a 2-sigma confidence interval, containing 95\% of the possible values, were they statistically distributed.

Finally, $\overline{u}$ has to be chosen to estimate the error in $X_i$. Most likely, $1\leq \overline{u} \leq U_{\mathrm{th}}$ but variation of $\overline{u}$ within that range does not alter the results strongly. Therefore, $\overline{u}=U_{\mathrm{th}}$ was chosen for simplicity.

A re-evaluation of the stellar neutron capture rates for all cases with $Z\geq10$
found in \citet{kadonis} was performed using the above procedure. Theoretical rates and MACS were calculated with the code SMARAGD, v0.8.4 \citep{smaragd,raureview}. The results are presented in Table \ref{tab:results}. Each line initially specifies the element symbol, charge $Z$, and mass number $A$ of the target nucleus. This is followed by 11 groups of values, for 11 temperatures $kT$: at 5, 10, 15, 20, 25, 30, 40, 50, 60, 80, and 100 keV. Each group contains, in this order, $X_{i^\mathrm{lab}}(T)$, $U_X(T)$, $R^*_\mathrm{new}(T)$, and the final uncertainty factor $U^*_\mathrm{new}(T)$ of $R^*_\mathrm{new}$ for the given temperature. These uncertainty factors $U^*_\mathrm{new}(T)$ also define the range within which a stellar rate should be varied in reaction sensitivity studies.

\begin{figure}
\includegraphics[angle=-90,width=\columnwidth]{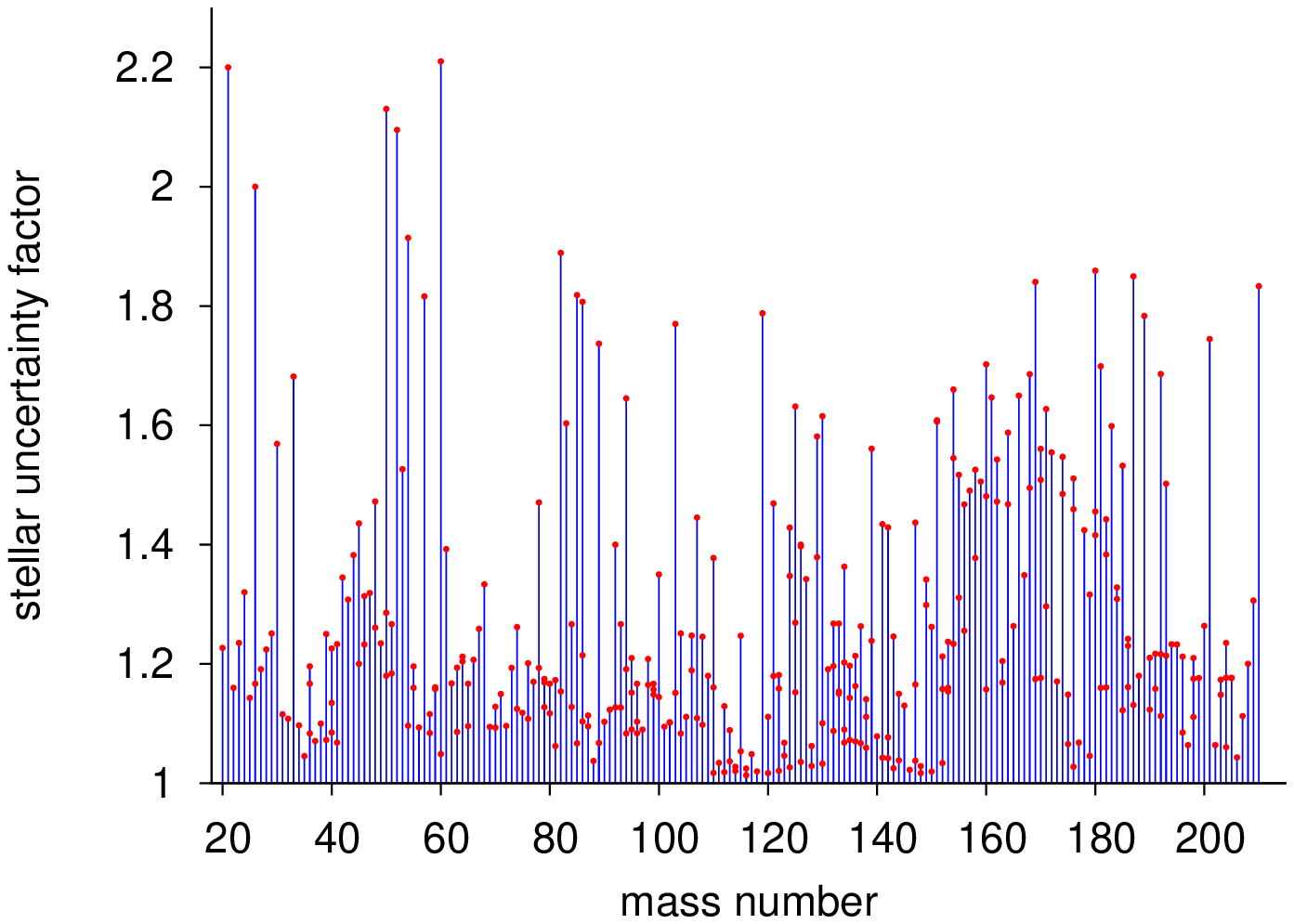}
\caption{Uncertainty factors $U^*_\mathrm{new}$ for stellar (n,$\gamma$) rates at $kT=30$ keV. (A color version of this figure is available in the online journal.)\label{fig:uncert}}
\end{figure}

As already pointed out in \citet{xfactor}, (n,$\gamma$)
reactivities in the higher mass part of the s-process are not well
constrained by experiment. Figure \ref{fig:uncert} shows the uncertainty factors $U^*_\mathrm{new}$ of the stellar reactivities at $kT=30$ keV. Again, large uncertainties are found in this
region already at that energy, despite of the fact that experimental uncertainties are small. The theoretical uncertainties are retained to a large extent.
This is the consequence of a small laboratory level contribution $X_{i^\mathrm{lab}}$ in target
nuclei with high nuclear level density, such as deformed nuclei.%
\footnote{There is one exception to this, the reaction $^{187}$Os(n,$\gamma$),
for which neutron transitions to excited states in $^{187}$Os were measured
separately recently \citep{mosc10,moscnscat,fujii10}. Because of this additional information,
the uncertainty on the stellar rate is similar to the one of the measurements.
The current version 0.3 of KADoNiS, however, does not include these
results.}
At higher temperature, the uncertainties approach those of the model for all nuclei because $X_{i^\mathrm{lab}}$ becomes small.
Even high precision measurements cannot improve the uncertainty unless they are able to investigate transitions on excited levels separately and use theory to construct the stellar rate from this information, applying Equations (\ref{eq:stellrate})--(\ref{eq:xfactor}).

An additional complication when performing renormalizations was mentioned in \S~\ref{sec:haufeshrenorm}: the question of reaction model applicability. This was ignored in the results presented
here but will be treated in more detail in the upcoming KADoNiS version.
It is expected, however, that the uncertainty stemming from the model
(in)applicability is anticorrelated with $X_i$, i.e., the statistical
Hauser-Feshbach model should be applicable when $X_i$ is small because
this implies a high level density.

Furthermore, it has to be pointed out that also values for $^{176}$Lu and $^{180}$Ta are included in Table \ref{tab:results}. It is not clear whether all levels in these nuclei are fully thermalized at the lower s-process temperatures and therefore these values may have a larger inherent error. Thermal population of all states is a fundamental assumption in the calculation of the stellar rates.

\section{Summary}

A general formalism to include experimental reaction cross sections into calculations of stellar rates was developed, which also allows to assess the reduction of uncertainties in the stellar rates by experiments. As an important example for the application of the procedure, stellar neutron capture reactivities from \cite{kadonis} were revised and the remaining uncertainties have been shown. Although the uncertainties in the stellar rates are close to the experimental values for a number of nuclei, many of the uncertainties remain considerably larger. This has important consequences for s-process models \citep[see, e.g.,][]{arl99} and the interpretation of meteoritic data \citep[see, e.g.,][]{qin,burk12} because it allows the rates of some reactions to vary within a larger range than previously assumed.

The revised reactivities will be included in the upcoming new version of KADoNiS.

\acknowledgments

This research was supported in part by the European Commission within the FP7 ENSAR/THEXO project and by the EuroGENESIS Collaborative Research Programme.

\clearpage

\begin{deluxetable}{rrrrrrrrrrrcrrrr}
\tabletypesize{\scriptsize}
\rotate
\tablecaption{Revised (n,$\gamma$) reactivities $R^*_\mathrm{new}$ and their uncertainty factors $U^*_\mathrm{new}$ for 11 plasma temperatures $T$; laboratory level contributions to the stellar rate $X_{i^\mathrm{lab}}$ and their uncertainty factors $U_X$ are also given. Negative values indicate that no information is available in \citet{kadonis} at that temperature.\label{tab:results}}
\tablewidth{0pt}
\tablehead{
&&& \multicolumn{4}{c}{$kT=5$ keV} & \multicolumn{4}{c}{$kT=10$ keV} & \dots & \multicolumn{4}{c}{$kT=100$ keV} \\
\colhead{Element} & \colhead{$Z$} & \colhead{$A$} & \colhead{$X$} & \colhead{$U_X$} & \colhead{$R^*_\mathrm{new}$} & \colhead{$U^*_\mathrm{new}$} & \colhead{$X$} & \colhead{$U_X$} & \colhead{$R^*_\mathrm{new}$} & \colhead{$U^*_\mathrm{new}$} &  \colhead{\dots} & \colhead{$X$} & \colhead{$U_X$} & \colhead{$R^*_\mathrm{new}$} & \colhead{$U^*_\mathrm{new}$}\\
&&&  &  & \colhead{(cm$^3$mol$^{-1}$s$^{-1}$)} & &  &  & \colhead{(cm$^3$mol$^{-1}$s$^{-1}$)} &  & \dots &  &  & \colhead{(cm$^3$mol$^{-1}$s$^{-1}$)} \\

}
\startdata
\dots &&&&&&&&&&&&&&& \\
W  & 74 & 182 & 1.0000 & 1.0000 & $3.7772\times 10^{7}$ & 1.0620 & 0.9997 & 1.0003 & $3.8496\times 10^7$ & 1.0626 & \dots & 0.2142 & 1.7858 & $6.8995\times 10^7$ & 1.8875 \\
W  & 74 & 183 & 0.9997 & 1.0003 & $9.4536\times 10^7$ & 1.0624 & 0.9754 & 1.0246 & $8.7088\times 10^7$ & 1.1069 & \dots & 0.1504 & 1.8496 & $8.9071\times 10^7$ & 1.9237 \\
W  & 74 & 184 & 1.0000 & 1.0000 & $3.5910\times 10^7$ & 1.0469 & 0.9999 & 1.0001 & $3.2889\times 10^7$ & 1.0471 & \dots & 0.2411 & 1.7589 & $4.4374\times 10^7$ & 1.8694 \\
W  & 74 & 185 & 0.9977 & 1.0023 & $1.1637\times 10^8$ & 1.2253 & 0.9746 & 1.0254 & $1.0062\times 10^8$ & 1.2603 & \dots & 0.2338 & 1.7662 & $5.1391\times 10^7$ & 1.8970 \\
W  & 74 & 186 & 1.0000 & 1.0000 & $3.8086\times 10^7$ & 1.0829 & 1.0000 & 1.0000 & $3.4755\times 10^7$ & 1.0879 & \dots & 0.2852 & 1.7148 & $2.9677\times 10^7$ & 1.8475 \\
Re & 75 & 185 & 1.0000 & 1.0000 & $2.5450\times 10^8$ & 1.0879 & 1.0000 & 1.0000 & $2.3289\times 10^8$ & 1.0879 & \dots & 0.6478 & 1.3522 & $2.1110\times 10^8$ & 1.5630 \\
Re & 75 & 186 & 1.0000 & 1.0000 & $1.8500\times 10^8$ & 1.1613 & 0.9945 & 1.0055 & $1.8000\times 10^8$ & 1.1613 & \dots & 0.1179 & 1.8821 & $2.8200\times 10^8$ & 1.1613 \\
\dots &&&&&&&&&&&&&&&
\enddata
\tablecomments{Table \ref{tab:results} is published in its entirety in the
electronic edition of the {\it Astrophysical Journal Letters}.  A portion is
shown here for guidance regarding its form and content.}
\end{deluxetable}

\end{document}



\title{Erratum: Formalism for inclusion of measured reaction cross sections in stellar rates including uncertainties
and its application to neutron capture in the s-process}

\author{Thomas Rauscher}
\affil{Department of Physics, University of Basel, CH-4056 Basel, Switzerland}

\begin{abstract}
The final rate uncertainty factors presented in \citet{formalism} have been calculated using an inappropriate value for the 
ground-state contribution in some cases. Revised rate uncertainty factors are presented here. Furthermore, the full version of Eq.\ 
(6), incomplete in \citet{formalism}, is given.
\end{abstract}


Revised low-energy neutron capture rates were given in \citet{formalism}, based on the identification of the ground-state 
contribution to stellar reaction rates. While the revised rates remain as given in \citet{formalism}, the assigned uncertainties 
were incorrectly calculated in some cases. In the calculation of the uncertainty factors $U^*_\mathrm{new}$ using Eq.\ (11) 
in \citet{formalism} the laboratory contributions $X_{i^\mathrm{lab}}$ \textit{before} the rate revision was used. Instead, the 
$X'_{i^\mathrm{lab}}$ of the new rate \textit{after} revision have to be used. As an alternative to using Eq.\ (11) in 
\citet{formalism} with the new $X'_{i^\mathrm{lab}}$ it is also possible to still use the previous $X_{i^\mathrm{lab}}$ but 
changing Eq.\ (11) in 
\citet{formalism} to
\begin{equation}
U^*_\mathrm{new}=U_{\mathrm{exp}}+\left(U_{\mathrm{th}}-U_{\mathrm{exp}}\right) \left(1-X_{i^\mathrm{lab}}
\right) \frac{r^*}{\left(1-X_{i^\mathrm{lab}}\right)r^*+r_{\mathrm{exp}}^{i^\mathrm{lab}}}\quad,\label{eq:uncertainty}
\end{equation}
where $r^*$ is the (theoretical) stellar rate with uncertainty factor $U_{\mathrm{th}}$, $r_{\mathrm{exp}}^{i^\mathrm{lab}}$ is the 
(measured) laboratory rate with uncertainty $U_{\mathrm{exp}}$, and $X_{i^\mathrm{lab}}$ is the (calculated) laboratory 
contribution to the stellar rate $r^*$ as obtained \textit{before} inclusion of any new information. Above Eq.\ 
(\ref{eq:uncertainty}) reduces to Eq.\ (11) of \citet{formalism} when the actual rate value did not change after inclusion of a new 
laboratory rate, e.g., when $r_{\mathrm{exp}}^{i^\mathrm{lab}}=r_{\mathrm{th}}^{i^\mathrm{lab}}$. Therefore, the old Eq.\ (11) of 
\citet{formalism} can still be used when only the uncertainties but not the absolute values of the rate changed.

\begin{figure}
\includegraphics[angle=-90,width=\columnwidth]{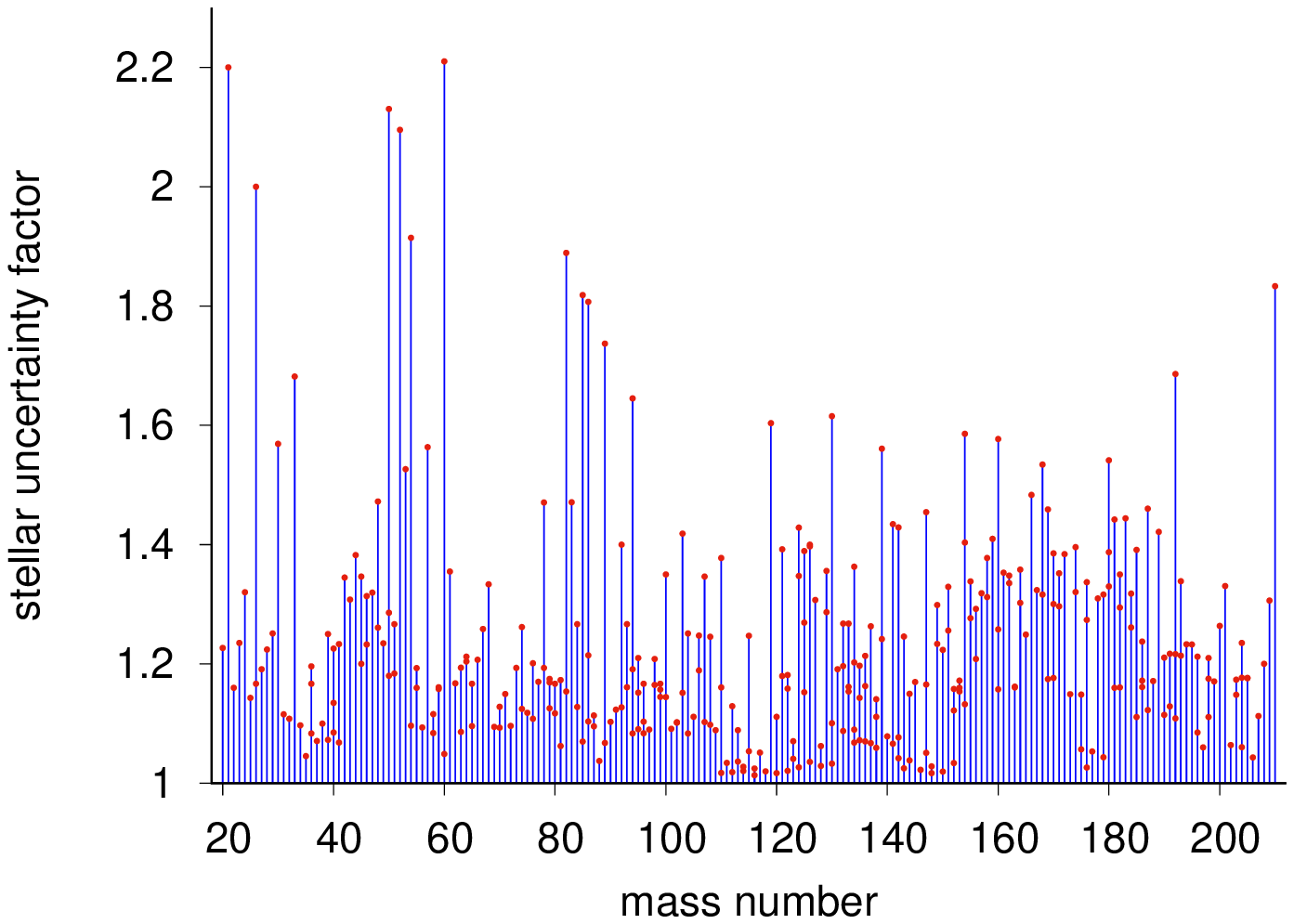}
\caption{Uncertainty factors $U^*_\mathrm{new}$ for stellar (n,$\gamma$) rates at $kT=30$ keV. (A color version of this figure is available in the online journal.)\label{fig:uncert}}
\end{figure}

Table \ref{tab:newuncert} gives the corrected values for the uncertainties $U^*_\mathrm{new}$. They supersede the values given in 
Table 1 of \citet{formalism}. Figure \ref{fig:uncert} provides an updated version of Fig.\ 1 in \citet{formalism}. It becomes 
apparent that the overall picture has not changed, with most uncertainty factors being larger than the experimental uncertainties 
due to the contribution of thermally excited states. As before, the largest uncertainties with $U^*_\mathrm{new}\gtrsim 1.8$ 
stem from the semi-empirical values given in \citet{kadonis}, which are based on theory. In comparison to the previously given 
uncertainties, most uncertainty reductions are found in the region around mass numbers $150\lesssim A \lesssim 190$ but they still 
remain larger than the experimental errors.

Finally, it has to be noted that Eq.\ (6) in \citet{formalism} is incomplete. The full expression for the effective cross section 
reads \citep{holadndt,raureview,sensi}
\begin{equation}
\sigma^\mathrm{eff}(E) = \sum_i \sum_j \frac{2J_i+1}{2J_0+1} \frac{E-E_i}{E}
\sigma^{i \rightarrow j}(E-E_i) 
\quad,
\label{eq:effcs}
\end{equation}
with $J_0$, $J_i$ being the spins of target ground state and target excited state $i$, and $\sigma^{i\rightarrow j}(E-E_i)$ being 
the partial cross section at energy $E-E_i$ from
target level $i$ with excitation energy $E_i$ to final level $j$ \citep[following][partial cross sections at zero or negative 
energies are set to zero]{fow74}.

\acknowledgments

Thanks go to Claudia Lederer-Woods for pointing out the need for amendment of the rate uncertainties as given in Table 1 of 
\citet{formalism}.

\clearpage

\begin{deluxetable}{rrrrrcr}
\tabletypesize{\scriptsize}
\rotate
\tablecaption{Revised uncertainty factors $U^*_\mathrm{new}$ of stellar (n,$\gamma$) reactivities at 11 plasma 
temperatures $T$.\label{tab:newuncert}}
\tablewidth{0pt}
\tablehead{
&&& \colhead{$kT=5$ keV} & \colhead{$kT=10$ keV} & \dots & \colhead{$kT=100$ keV} \\
\colhead{Element} & \colhead{$Z$} & \colhead{$A$} & \colhead{$U^*_\mathrm{new}$}  & 
\colhead{$U^*_\mathrm{new}$} & \colhead{\dots} & \colhead{$U^*_\mathrm{new}$}\\
}
\startdata
\dots &&&&& \\
W  & 74 & 182 & 1.0620 & 1.0627 & \dots & 1.6290 \\
W  & 74 & 183 & 1.0625 & 1.1091 & \dots & 1.5898 \\
W  & 74 & 184 & 1.0469 & 1.0471 & \dots & 1.5772 \\
W  & 74 & 185 & 1.2250 & 1.2525 & \dots & 1.5202 \\
W  & 74 & 186 & 1.0829 & 1.0829 & \dots & 1.4043 \\
Re & 75 & 185 & 1.0879 & 1.0879 & \dots & 1.3123 \\
Re & 75 & 186 & 1.1613 & 1.1613 & \dots & 1.1613 \\
\dots &&&&&
\enddata
\tablecomments{Table \ref{tab:newuncert} is published in its entirety in the
electronic edition of the {\it Astrophysical Journal Letters}.  A portion is
shown here for guidance regarding its form and content.}
\end{deluxetable}